\documentstyle[12pt]{article}
\newlength{\extraspace}
\newlength{\extraspaces}
\newcommand{\be}{\begin{equation}
\addtolength{\abovedisplayskip}{\extraspaces}
\addtolength{\belowdisplayskip}{\extraspaces}
\addtolength{\abovedisplayshortskip}{\extraspace}
\addtolength{\belowdisplayshortskip}{\extraspace}}
\newcommand{\ee}{\end{equation}}
\newcommand{\ba}{\begin{eqnarray}
\addtolength{\abovedisplayskip}{\extraspaces}
\addtolength{\belowdisplayskip}{\extraspaces}
\addtolength{\abovedisplayshortskip}{\extraspace}
\addtolength{\belowdisplayshortskip}{\extraspace}}
\newcommand{\ea}{\end{eqnarray}}
\newcommand{\nonu}{\nonumber \\[.5mm]}
\newcommand{\A}{&\!\!\!}
\begin{document}
\thispagestyle{empty}
\begin{flushright}
SIT-LP-06/06 \\
{\tt hep-th/0606169} \\
June, 2006
\end{flushright}
\vspace{7mm}
\begin{center}
{\large \bf Linearizing $N = 3$ nonlinear supersymmetry \\[2mm]
in two dimensions 
} \\[20mm]
{\sc Kazunari Shima}
\footnote{
\tt e-mail: shima@sit.ac.jp} \ 
and \ 
{\sc Motomu Tsuda}
\footnote{
\tt e-mail: tsuda@sit.ac.jp} 
\\[5mm]
{\it Laboratory of Physics, 
Saitama Institute of Technology \\
Fukaya, Saitama 369-0293, Japan} \\[20mm]
\begin{abstract}
We investigate for $N = 3$ supersymmetry (SUSY) in $D = 2$ 
the algebraic relation between the Volkov-Akulov (VA) model 
of nonlinear (NL) SUSY 
and a (renormalizable) SO(3) vector supermultiplet of linear (L) SUSY. 
We derive SUSY and SO(3) invariant relations between component fields 
of the vector supermultiplet 
and Nambu-Goldstone (NG) fermions of the VA model at leading orders 
by using three arbitrary dimensionless parameters 
which can be recasted as the vacuum expectation values 
of auxiliary fields in the vector supermultiplet. 
Two different irreducible representations of SO(3) super-Poincar\'e symmetry 
which appear in the same massless state 
are compatible with each other in the linearization of NL SUSY. 
The equivalence of a NL SUSY VA action to a free L SUSY action 
containing the Fayet-Iliopoulos (FI) $D$ term 
which indicates a spontaneously SUSY breaking 
is also discussed explicitly according to the SUSY invariant relations. 
\end{abstract}
\end{center}

\newpage

Nonlinear supersymmetry (NL SUSY) \cite{VA} is realized 
by means of Nambu- Goldstone (NG) fermions \cite{SS}-\cite{O} 
which correspond to a spontaneously SUSY breaking (SSB). 
It is (globally) extended into the curved spacetime 
by defining a tangent spacetime which posesses 
the NG fermion degrees of freedom as SL(2,C) Grassman coordinates 
besides the Minkowski one \cite{KS1}. 
This extension shedded light on the way 
towards a composite unified theory of spacetime and matter 
based upon SO(10) super-Poincar\'e (SP) algebra, 
i.e., the superon-graviton model (SGM) \cite{KS1,KS2}, 
which was constructed as a NL SUSY general relativity 
\cite{KS1,ST1,ST2,ST3}. 
In order to investigate the low energy physical contents 
for the basic NL SUSY Einstein-Hilbert type action of the SGM, 
it is important to know the relation between the NL 
and linear (L) SUSY in detail. 
In the flat spacetime case, it has been known 
that the NL SUSY model of Volkov-Akulov (VA) \cite{VA} 
describes various (renormalizable) L supermultiplets 
\cite{WZ1}-\cite{Fa} with the Fayet-Iliopoulos (FI) $D$ term 
indicating the SSB through a linearization 
at least for $N = 1$ and $N = 2$ SUSY. 
Indeed, for $N = 1$ SUSY the VA model describes a scalar supermultiplet 
\cite{IK}-\cite{UZ}, while it also expresses a U(1) axial vector one 
\cite{IK,STT1}. 
For $N = 2$ SUSY the VA model is (algebraically) equivalent 
to a SU(2) $\times$ U(1) vector supermultiplet \cite{STT2}. 

The linearization of $N = 1$ and $N = 2$ NL SUSY in flat spacetime 
shows that the VA model has richer physical 
structure for higher $N$ SUSY. 
However, the detailed relation of the NL and the L SUSY 
for higher $N$ is unknown so far. 
Therefore, as a preliminary to know this, 
we focus in this letter on $N = 3$ SUSY 
and investigate the relation between the VA model and a L SUSY one 
which has the multiplet structure as shown below in Eq.(\ref{irrep}). 
For $N = 3$ SUSY, the tower of the helicity states 
for the irreducible representation of SO(3) SP algebra is 
\be
[ \ \underline{1} (+1), 
\underline{3} \left( +{1 \over 2} \right), 
\underline{3} (0), 
\underline{1} \left( -{1 \over 2} \right) \ ] 
+ [\ {\rm CPT\ conjugate}\ ], 
\label{irrep}
\ee
where $\underline{n} (\lambda)$ means the dimension $\underline{n}$ 
and the helicity $\lambda$ of the irreducible representation. 
Note that two different irreducible representations of SO(3) SP symmetry 
appear in the same massless state, 
which is a characteristic aspect for higher $N$ SUSY. 

In the linearization of NL SUSY, 
it is important to find SUSY invariant relations 
which express fundamental fields of L supermultiplets 
as composites of NG fermions 
and which reproduce L SUSY transformations among the fundamental fields 
under NL SUSY transformations of NG fermions. 
The SUSY invariant relations are obtained by means of systematic procedure 
\cite{IK}-\cite{STT1} in the superspace formalism 
or by using heuristic arguments \cite{R,STT2} 
in which the component fields of L SUSY is directly expanded 
in terms of the NG fermions in a SUSY invariant way. 
In this letter we rely on the heuristic method 
and derive the SUSY invariant relations for $N = 3$ SUSY 
at leading orders by using $D = 2$ calculations for simplicity. 
In particular, we show explicitly that 
the two different irreducible representations in the same massless state 
of Eq.(\ref{irrep}) are compatible with each other 
in the SUSY and SO(3) invariant linearization of NL SUSY. 
According to the SUSY invariant relations 
we pass from a free L SUSY action with the FI $D$ term 
to the NL SUSY VA action. 

Let us introduce the NL realization of SUSY \cite{VA} 
for arbitrary $N$ \cite{BV} in $D = 2$. 
\footnote{
Minkowski spacetime indices are denoted by $a, b, \cdots = 0, 1$ 
and SO(N) internal indices are $i, j, \cdots = 1, 2, \cdots, N$. 
The Minkowski spacetime metric is 
${1 \over 2}\{ \gamma^a, \gamma^b \} = \eta^{ab} = {\rm diag}(+, -)$ 
and $\sigma^{ab} = {i \over 2}[\gamma^a, \gamma^b] 
= i \epsilon^{ab} \gamma_5$ $(\epsilon^{01} = 1 = - \epsilon_{01})$, 
where we use the $\gamma$ matrices defined as $\gamma^0 = \sigma^2$, 
$\gamma^1 = i \sigma^1$, $\gamma_5 = \gamma^0 \gamma^1 = \sigma^3$ 
with $\sigma^I (I = 1, 2, 3)$ being Pauli matrices. 
(As for the conventions in $D = 2$, for example see \cite{UZ}.)
}
NL SUSY transformations of (Majorana) NG fermions $\psi^i$ 
are parametrized by constant (Majorana) spinor parameters $\zeta^i$, 
\be
\delta \psi^i = {1 \over \kappa} \zeta^i 
- i \kappa \bar\zeta^j \gamma^a \psi^j \partial_a \psi^i, 
\label{NLSUSY}
\ee
which correspond to supertranslations of the $\psi^i$ 
and the Minkowski coordinate $x^a$. 
In Eq.(\ref{NLSUSY}) $\kappa$ is a constant whose dimension is $({\rm mass})^{-1}$. 
Eq.(\ref{NLSUSY}) satisfies the following closed off-shell commutator algebra, 
\be
[ \delta_{Q1}, \delta_{Q2} ] = \delta_P(\Xi^a), 
\label{N3D2com}
\ee
where $\delta_P(\Xi^a)$ means a translation with a generator 
$\Xi^a = - 2 \bar\zeta_1^i \gamma^a \zeta_2^i$. 

An action, which is invariant under the NL SUSY transformation (\ref{NLSUSY}), 
is constructed based upon a NL SUSY invariant differential one-form defined as 
\ba
\omega^a \A \A 
= d x^a - i \kappa^2 \bar\psi^i \gamma^a d \psi^i 
\nonu
\A \A = (\delta^a_b 
- i \kappa^2 \bar\psi^i \gamma^a \partial_b \psi^i) \ dx^b 
\nonu
\A \A = (\delta^a_b + t^a{}_b) \ dx^b 
\nonu
\A \A = w^a{}_b \ dx^b. 
\label{one-form}
\ea
From Eq.(\ref{one-form}) the NL SUSY action is given 
as the volume form in $D = 2$, 
\ba
S_{\rm NL} = \A \A - {1 \over {2 \kappa^2}} \int \omega^0 \wedge \omega^1 
\nonu
= \A \A - {1 \over {2 \kappa^2}} \int d^2 x \ \vert w \vert 
\nonu
= \A \A - {1 \over {2 \kappa^2}} \int d^2 x 
\left\{ 1 + t^a{}_a + {1 \over 2!}(t^a{}_a t^b{}_b - t^a{}_b t^b{}_a) 
\right\} 
\nonu
= \A \A - {1 \over {2 \kappa^2}} \int d^2 x 
\left\{ 1 - i \kappa^2 \bar\psi^i \!\!\not\!\partial \psi^i 
- {1 \over 2} \kappa^4 
( \bar\psi^i \!\!\not\!\partial \psi^i \bar\psi^j \!\!\not\!\partial \psi^j 
- \bar\psi^i \gamma^a \partial_b \psi^i \bar\psi^j \gamma^b \partial_a \psi^j ) 
\right\} 
\nonu
= \A \A - {1 \over {2 \kappa^2}} \int d^2 x 
\left\{ 1 - i \kappa^2 \bar\psi^i \!\!\not\!\partial \psi^i 
\right. 
\nonu
\A \A 
\left. 
- {1 \over 2} \kappa^4 \epsilon^{ab} 
( \bar\psi^i \psi^j \partial_a \bar\psi^i \gamma_5 \partial_b \psi^j 
+ \bar\psi^i \gamma_5 \psi^j \partial_a \bar\psi^i \partial_b \psi^j ) 
\right\}, 
\label{NLSUSYact}
\ea
where the second term, 
$-{1 \over {2 \kappa^2}} t^a{}_a = {i \over 2} 
\bar\psi^i \!\!\not\!\partial \psi^i$, 
is the kinetic term for $\psi^i$. 

In order to linearize the above NL SUSY model by focusing on $N = 3$ SUSY, 
we consider a $N = 3$ L supermultiplet in $D = 2$ 
which has the multiplet structure (\ref{irrep}). 
We denote component fields of the off-shell massless L supermultiplet 
as $v^a$ for a vector field, 
$\lambda^i$ for triplet (Majorana) fermions, 
$A^i$ for triplet (real) scalar fields 
and $\chi$ for a singlet (Majorana) fermion 
in addition to $\phi$ for a scalar field 
and $F^i$ for three auxiliary scalar fields. 
L SUSY transformations of those component fields generated by $\zeta^i$ 
are defined as 
\ba
\A \A 
\delta_\zeta v^a 
= i \bar\zeta^i \gamma^a \lambda^i, 
\label{LSUSY-v}
\\
\A \A 
\delta_\zeta \lambda^i 
= \epsilon^{ijk} ( F^j - i \!\!\not\!\partial A^j ) \zeta^k 
+ {1 \over 2} \epsilon^{ab} F_{ab} \gamma_5 \zeta^i 
- i \gamma_5 \!\!\not\!\partial \phi \zeta^i, 
\\
\A \A 
\delta_\zeta A^i 
= \epsilon^{ijk} \bar\zeta^j \lambda^k - \bar\zeta^i \chi, 
\\
\A \A 
\delta_\zeta \chi 
= ( F^i + i \!\!\not\!\partial A^i ) \zeta^i, 
\\
\A \A 
\delta_\zeta \phi 
= \bar\zeta^i \gamma_5 \lambda^i, 
\\
\A \A 
\delta_\zeta F^i 
= - i \epsilon^{ijk} \bar\zeta^j \!\!\not\!\partial \lambda^k 
- i \bar\zeta^i \!\!\not\!\partial \chi, 
\label{LSUSY-F}
\ea
where $F_{ab} = \partial_a v_b - \partial_b v_a$. 
These transformations (\ref{LSUSY-v}) to (\ref{LSUSY-F}) 
satify the closed off-shell commutator algebra 
with a U(1) gauge transformation of $v^a$, 
\be
[ \delta_{Q1}, \delta_{Q2} ] = \delta_P(\Xi^a) + \delta_g(\theta), 
\label{N3D2com-gauge}
\ee
where $\delta_g(\theta)$ is the U(1) gauge transformation 
with a generator $\theta = 2 (i \bar\zeta_1^i \gamma^a \zeta_2^i v_a 
- \epsilon^{ijk} \bar\zeta_1^i \zeta_2^j A^k 
- \bar\zeta_1^i \gamma_5 \zeta_2^i \phi)$. 

Let us show SUSY invariant relations 
between the component fields 
($v^a$, $\lambda^i$, $A^i$, $\chi$, $\phi$, $F^i$) 
and the NG fermions $\psi^i$ at the leading orders. 
From the experiences in $N = 1$ and $N = 2$ SUSY \cite{IK}-\cite{STT2}, 
we suppose 
\ba
\lambda^i \A = \A \epsilon^{ijk} \xi^j \psi^k + {\cal O}(\kappa^2), 
\nonu
\chi \A = \A \xi^i \psi^i + {\cal O}(\kappa^2), 
\nonu
F^i \A = \A {1 \over \kappa} \xi^i + {\cal O}(\kappa), 
\nonu
({\rm other\ fields}) \A = \A {\cal O}(\kappa), 
\label{leading}
\ea
where $\xi^i$ are three arbitrary real parameters 
satisfying $(\xi^i)^2 = 1$, 
and we expect the broken L SUSY with 
the vacuum expectation values (vev) $F^i = (1/\kappa) \xi^i$ 
derived from a free L SUSY action as shown later. 
Higher order terms in Eq.(\ref{leading}) are obtained 
such that the NL SUSY tranformations (\ref{NLSUSY}) 
reproduce the L SUSY ones (\ref{LSUSY-v}) to (\ref{LSUSY-F}). 

After some calculations 
($v^a$, $\lambda^i$, $A^i$, $\chi$, $\phi$, $F^i$) 
are expanded in terms of $\psi^i$ 
in the SUSY and SO(3) invariant way as 
\ba
v^a 
\A = \A - {i \over 2} \kappa \epsilon^{ijk} \xi^i 
\bar\psi^j \gamma^a \psi^k 
( 1 - i \kappa^2 \bar\psi^l \!\!\not\!\partial \psi^l ) 
+ {1 \over 4} \kappa^3 
\epsilon^{ab} \epsilon^{ijk} \xi^i \partial_b 
( \bar\psi^j \gamma_5 \psi^k \bar\psi^l \psi^l ) 
+ {\cal O}(\kappa^5), 
\label{expand-v}
\\
\lambda^i 
\A = \A \epsilon^{ijk} \xi^j \psi^k 
( 1 - i \kappa^2 \bar\psi^l \!\!\not\!\partial \psi^l ) 
\nonu
\A \A 
+ {i \over 2} \kappa^2 \xi^j \partial_a 
\{ \epsilon^{ijk} \gamma^a \psi^k \bar\psi^l \psi^l 
+ \epsilon^{ab} \epsilon^{jkl} 
( \gamma_b \psi^i \bar\psi^k \gamma_5 \psi^l 
- \gamma_5 \psi^i \bar\psi^k \gamma_b \psi^l ) \} 
+ {\cal O}(\kappa^4), 
\\
A^i 
\A = \A \kappa \left( {1 \over 2} \xi^i \bar\psi^j \psi^j 
- \xi^j \bar\psi^i \psi^j \right) 
( 1 - i \kappa^2 \bar\psi^k \!\!\not\!\partial \psi^k ) 
- {i \over 2} \kappa^3 \xi^j \partial_a 
( \bar\psi^i \gamma^a \psi^j \bar\psi^k \psi^k ) 
+ {\cal O}(\kappa^5), 
\\
\chi 
\A = \A \xi^i \psi^i 
( 1 - i \kappa^2 \bar\psi^j \!\!\not\!\partial \psi^j ) 
+ {i \over 2} \kappa^2 \xi^i \partial_a 
( \gamma^a \psi^i \bar\psi^j \psi^j ) 
+ {\cal O}(\kappa^4), 
\\
\phi 
\A = \A - {1 \over 2} \kappa \epsilon^{ijk} \xi^i \bar\psi^j \gamma_5 \psi^k 
( 1 - i \kappa^2 \bar\psi^l \!\!\not\!\partial \psi^l ) 
- {i \over 4} \kappa^3 
\epsilon^{ab} \epsilon^{ijk} \xi^i \partial_a 
( \bar\psi^j \gamma_b \psi^k \bar\psi^l \psi^l ) 
+ {\cal O}(\kappa^5), 
\\
F^i 
\A = \A {1 \over \kappa} \xi^i 
\left\{ 1 - i \kappa^2 \bar\psi^j \!\!\not\!\partial \psi^j 
- {1 \over 2} \kappa^4 
( \bar\psi^j \!\!\not\!\partial \psi^j \bar\psi^k \!\!\not\!\partial \psi^k 
- \bar\psi^j \gamma^a \partial_b \psi^j \bar\psi^k \gamma^b \partial_a \psi^k ) 
\right\} 
\nonu
\A \A 
- i \kappa \xi^j \partial_a \{ \bar\psi^i \gamma^a \psi^j 
( 1 - i \kappa^2 \bar\psi^k \!\!\not\!\partial \psi^k ) \} 
- {1 \over 8} \kappa^3 
\Box 
\{ ( \xi^i \bar\psi^j \psi^j - 4 \xi^j \bar\psi^i \psi^j ) 
\bar\psi^k \psi^k \} 
\nonu
\A \A 
+ {\cal O}(\kappa^5) 
\nonu
\A = \A 
{1 \over \kappa} \xi^i \vert w \vert 
- i \kappa \xi^j \partial_a \{ \bar\psi^i \gamma^a \psi^j 
( 1 - i \kappa^2 \bar\psi^k \!\!\not\!\partial \psi^k ) \} 
- {1 \over 8} \kappa^3 
\Box 
\{ ( \xi^i \bar\psi^j \psi^j - 4 \xi^j \bar\psi^i \psi^j ) 
\bar\psi^k \psi^k \} 
\nonu
\A \A 
+ {\cal O}(\kappa^5). 
\label{expand-F}
\ea
Note that the terms proportional to 
$(1 - i \kappa^2 \bar\psi^i \!\!\not\!\partial \psi^i)$ 
appear in Eqs. from (\ref{expand-v}) to (\ref{expand-F}). 
We expect that those terms are calculated as the parts of $\vert w \vert$. 
Such systematics of $D = 2$ SUSY invariant relations 
(for $N = 1$ SUSY, for example see \cite{R}) will be confirmed 
by continuing the above calculations 
in Eqs. from (\ref{expand-v}) to (\ref{expand-F}) 
at higher orders of $\kappa$ 
or by means of the systematic arguments in the superspace formalism. 

The transformation of Eq.(\ref{expand-v}) 
with respect to $\psi^i$ under Eq.(\ref{NLSUSY}) 
gives the U(1) gauge transformation 
besides the L SUSY one as 
\be
\delta_\zeta v^a(\psi) 
= i \bar\zeta^i \gamma^a \lambda^i(\psi) + \partial^a X(\zeta; \psi), 
\label{NLSUSY-v}
\ee
where $X(\zeta; \psi)$ is the U(1) gauge transformation parameter 
defined by 
\be
X(\zeta; \psi) 
= - {1 \over 2} \kappa^2 \epsilon^{ijk} \xi^i 
(\bar\zeta^j \psi^k \bar\psi^l \psi^l 
+ \bar\psi^j \gamma_5 \psi^k \bar\zeta^l \gamma_5 \psi^l 
+ \bar\psi^j \gamma_a \psi^k \bar\zeta^l \gamma^a \psi^l). 
\label{gauge-parameter}
\ee
Gauge invariant quantities like $F_{ab}(\psi)$ 
transform exactly same as the L SUSY transformation. 
Since Eq.(\ref{gauge-parameter}) satisfies 
\be
\delta_{\zeta_2} X(\zeta_1; \psi) 
- \delta_{\zeta_2} X(\zeta_1; \psi) 
= - \theta(\zeta_1, \zeta_2; A^i(\psi), v^a(\psi), \phi(\psi)), 
\ee
the commutator for $v^a(\psi)$ through Eq.(\ref{NLSUSY-v}) 
does not contain the U(1) gauge transformation term $\delta_g(\theta)$. 
This is the same case as in the relation between the VA model 
and the SU(2) $\times$ U(1) vector supermutiplet 
for $N = 2$ SUSY in $D = 4$ \cite{STT2}, 
i.e., it should be the case that the commutator on $\psi^i$ 
does not contain the $\delta_g(\theta)$. 

The form of NL and L SUSY actions is not used 
in the derivation of the SUSY invariant relations 
(\ref{expand-v}) to (\ref{expand-F}). 
By using those relations, we discuss the equivalence 
of the NL SUSY action (\ref{NLSUSYact}) 
for $N = 3$ SUSY with the following free L SUSY action 
for ($v^a$, $\lambda^i$, $A^i$, $\chi$, $\phi$, $F^i$) 
defined as 
\ba
S_{\rm L} = \A \A \int d^2 x \left\{ 
{1 \over 2} (\partial_a A^i)^2 - {1 \over 4} (F_{ab})^2 
+ {i \over 2} \bar\lambda^i \!\!\not\!\partial \lambda^i 
+ {i \over 2} \bar\chi \!\!\not\!\partial \chi 
+ {1 \over 2} (\partial_a \phi)^2 
\right. 
\nonu
\A \A 
\left. + {1 \over 2} (F^i)^2 
- {1 \over \kappa} \xi^i F^i 
\right\}. 
\label{LSUSYact}
\ea
Note that the last term proportional to $\kappa^{-1}$ 
in Eq.(\ref{LSUSYact}) is an analog of the FI $D$ term 
and the field equations for the auxiliary fields 
give the vev $F^i = {1 \over \kappa} \xi^i$ indicating the SSB. 
By substituting Eqs. from (\ref{expand-v}) to (\ref{expand-F}) 
into the action (\ref{LSUSYact}), 
we can show 
\be
S_{\rm L} = S_{\rm NL} + [ \ {\rm surface\ terms} \ ] 
\ee
at least up to ${\cal O}(\kappa^2)$. 

To summarize our results, 
we have investigated for $N = 3$ SUSY in $D = 2$ 
the relation between the VA model 
and the (renormalizable) SO(3) vector supermultiplet 
which corresponds to Eq.(\ref{irrep}). 
By means of the heuristic arguments, 
the SUSY and SO(3) invariant relations 
have been derived up to ${\cal O}(\kappa^2)$ or ${\cal O}(\kappa^3)$ 
in Eqs. from (\ref{expand-v}) to (\ref{expand-F}), 
in which we have confirmed that $\lambda^i$ and $\chi$, 
i.e., the two different irreducible representations 
in the same massless state of Eq.(\ref{irrep}) 
are compatible with each other in the linearization of NL SUSY. 
We have calculated the commutator for $v^a(\psi)$ of Eq.(\ref{expand-v}) 
by using the NL SUSY transformations (\ref{NLSUSY}) of $\psi^i$, 
and we have shown that it does not contain the U(1) gauge transformation term 
as in the linearization of the $N = 2$ NL SUSY in $D = 4$ \cite{STT2}. 
The equivalence of the NL SUSY action (\ref{NLSUSYact}) 
with the free L SUSY action (\ref{LSUSYact}) 
through the relations (\ref{expand-v}) to (\ref{expand-F}) 
has been also shown explicitly at least up to ${\cal O}(\kappa^2)$. 
The sytematic arguments in the superspace formalism may be useful 
in the derivation of the SUSY invariant relations at all orders 
of $\kappa$ and in order to discuss some systematics 
in the $D = 2$ relations.

%
%

\newpage

%
\newcommand{\NP}[1]{{\it Nucl.\ Phys.\ }{\bf #1}}
\newcommand{\PL}[1]{{\it Phys.\ Lett.\ }{\bf #1}}
\newcommand{\CMP}[1]{{\it Commun.\ Math.\ Phys.\ }{\bf #1}}
\newcommand{\MPL}[1]{{\it Mod.\ Phys.\ Lett.\ }{\bf #1}}
\newcommand{\IJMP}[1]{{\it Int.\ J. Mod.\ Phys.\ }{\bf #1}}
\newcommand{\PR}[1]{{\it Phys.\ Rev.\ }{\bf #1}}
\newcommand{\PRL}[1]{{\it Phys.\ Rev.\ Lett.\ }{\bf #1}}
\newcommand{\PTP}[1]{{\it Prog.\ Theor.\ Phys.\ }{\bf #1}}
\newcommand{\PTPS}[1]{{\it Prog.\ Theor.\ Phys.\ Suppl.\ }{\bf #1}}
\newcommand{\AP}[1]{{\it Ann.\ Phys.\ }{\bf #1}}

\end{document}